\documentclass[conference]{IEEEtran}
\IEEEoverridecommandlockouts
\usepackage{cite}
\usepackage{amsmath,amssymb,amsfonts}
\usepackage{graphicx}
\usepackage{subcaption}
\usepackage{textcomp}
\usepackage{xcolor}
\usepackage{booktabs}
\usepackage{multirow}
\usepackage{tabularx}
\usepackage{fancyhdr}
\usepackage[ruled,vlined]{algorithm2e}
\include{pythonlisting}

\begin{document}

\title{DPNCT: A Differential Private Noise Cancellation Scheme for Load Monitoring and Billing for Smart Meters \\
}

%\author{\IEEEauthorblockN{Khadija Hafeez}
%\IEEEauthorblockA{\textit{Department of Computer Science} \\
%\textit{Munster Technological University (MTU),}\\
%Ireland \\
%khadija.hafeez@mycit.ie}
%\and
%\IEEEauthorblockN{Mubashir Husain Rehmani}
%\IEEEauthorblockA{\textit{Department of Computer Science} \\
%\textit{Munster Technological University (MTU),}\\
%Ireland \\
%mubashir.rehmani@cit.ie}
%\and
%\IEEEauthorblockN{Donna O'Shea}
%\IEEEauthorblockA{\textit{Department of Computer Science} \\
%\textit{Munster Technological University (MTU),}\\
%Ireland \\
%donna.oshea@cit.ie}
%
%}

\author{\IEEEauthorblockN{Khadija Hafeez\IEEEauthorrefmark{1}, Mubashir Husain Rehmani\IEEEauthorrefmark{1}, Donna O'Shea\IEEEauthorrefmark{1}\\}
\IEEEauthorblockA{\IEEEauthorrefmark{1} Munster Technological University (MTU), Cork, Ireland%\thanks{This publication has emanated from research conducted with the financial support of Science Foundation Ireland (SFI) and is funded under the Grant Number 18/CRT/6222.}
}
}

\maketitle

\begin{abstract}
Highly accurate profiles of consumers daily energy usage are reported to power grid via smart meters which enables smart grid to effectively regulate power demand and supply. However, consumer's energy consumption pattern can reveal personal and sensitive information regarding their lifestyle. Therefore, to ensure users privacy, differentially distributed noise is added to the original data. This technique comes with a trade off between privacy of the consumer versus utility of the data in terms of providing services like billing, Demand Response schemes, and Load Monitoring. In this paper, we propose a technique - Differential Privacy with Noise Cancellation Technique (DPNCT) - to maximize utility in aggregated load monitoring and fair billing while preserving users' privacy by using noise cancellation mechanism on differentially private data. We introduce noise to the sensitive data stream before it leaves smart meters in order to guarantee privacy at individual level. Further, we evaluate the effects of different periodic noise cancelling schemes on privacy and utility i.e., billing and load monitoring. Our proposed scheme outperforms the existing scheme in terms of preserving the privacy while accurately calculating the bill. %hourly noise cancelling scheme provides the least $5\%$ Mean Absolute Error (MAE) in total energy consumption and $6\%$ in total bill as compared to $100\%$ MAE in Differentially Private Dynamic Pricing for Demand Response (DRDP) \cite{Hassan2020} masked total energy consumption and $70\%$ in total bill using DRDP masked data. 
% \textcolor{blue}{KHADIJA, kindly update this blue color text! We concluded that hourly noise cancelling scheme provides the least $5\%$ Mean Absolute Error (MAE) in total energy consumption.   }
\end{abstract}


\section{Introduction}

The term Cyber Physical System (CPS) refers to large scale intelligent, reactive and highly configurable hybrid system which has both physical and computational properties. In smart grids, CPS is enabled through smart meters, which are entities that collect end user consumption data at high frequency in real time, transmitting this data to the utility grid provider. Such real time collection of end-user data facilitates Demand Response (DR) schemes which influence the customer demand of energy usage from peak time to off peak time for better distribution and generation of load. Such DR schemes and detailed collection of energy usage data can reveal sensitive and private information regarding consumer's life style \cite{inferPrivLeak}.

%Smart Grids technology is a tightly coupled CPS that integrates physical systems like system power network infrastructure and cyber systems such as sensors, where two way communication between consumer and power grid enables economical and efficient energy consumption using Demand Response (DR) schemes. DR schemes influence the customer demand of energy usage from peak time to off peak time for better distribution and generation of load. Smart meters, a key entity in the smart grid system, transmit information of the end-user electricity load consumption at a high frequency in real time to optimise DR schemes, however, the detailed collection of energy usage data can reveal sensitive and private information regarding consumer's life style \cite{inferPrivLeak}. 
% Recent research\cite{inferPrivLeak} pointed out the extent of privacy invasion that can happen using consumer's energy consumption data. 
Molina-Markham et al. \cite{privInvationUmass} shows that the power consumption pattern can reveal personal information including the time periods when the consumer is not at home, the type of electrical devices that are being used at a household, and any change in the habits of the consumer such as sleeping and eating. This information can be used for targeted marketing and can pose a serious security threat to the consumer.\par
%An aggregator is an intermediary between smart meter and power grid, which collects the smart meter data at a network level and provides services, including but not limited to bill calculation of individuals, load monitoring, and enforcement of DR schemes.
In order to address the challenge of privacy invasion, Differential Privacy (DP) first proposed by Dwork et al. \cite{dwork2014algorithmic}, which adds noise to the critical data in a way that addition, deletion or change in an individual record makes insignificant difference to the overall output. A central architectural component of DP is an aggregator which acts as an intermediary between smart meter and power grid, which collects the smart meter data at a network level and provides services, including but not limited to bill calculation of individuals, load monitoring, and enforcement of DR schemes~\cite{Hassan2020Comst}. The goal of using DP for smart meter data is to release the statistics to the aggregator for critical decision making in DR schemes while preserving user's privacy. The challenge associated with this goal is how to provide a mechanism that preserves individual user privacy, enabling the aggregator to calculate total energy consumption of all smart meters in an area at an instant in time $t$ and individual users over a period of time $T$.

\begin{table*}[ht]
\centering
\vspace{0.25cm}
\caption{Comparison of Techniques for Privacy Preserving using Differential Privacy in smart meters}

\begin{tabularx}{\textwidth}{|c|m{10em}|m{5em}|X|m{15em}|}
\hline
\textbf{Ref. No}& \textbf{\textit{Focus}}& \textbf{\textit{Privacy Type}}& \textbf{\textit{Working Mechanism }}& \textbf{\textit{Limitation}}\\
\hline

\cite{dream}& 
Differential Privacy without trusting third party&  
Differential Privacy with Encryption& 
Multiple exchange of encrypted messages with aggregator for differentially private data& 
Partial fault tolerance, 
Increased utilization of bandwidth,
Privacy for aggregated data only
\\
\hline

\cite{Eibl2017}& 
Infinite Divisibility of Laplacian Noise with post processing smoothing&
Differential Privacy& 
Adding gamma distributed noise to each individual agent using infinite divisible laplace distribution&
Privacy for Aggregated information only\\

\hline
\cite{Hassan2020,Hassan2021}& 
Dynamic Pricing and Privacy&  
Differential Privacy& 
Dual Differential Privacy with Dynamic pricing using trusted third party& 
Too much trust on third party for storing real data and calculation of bills, No analysis on the usability of differentially private data at grid level\\

\hline
\cite{Barbosa2016}& 
Privacy for Appliance Usage&  
Differential Privacy& 
Differential privacy using Laplacian noise with filtering attack analysis to preserve appliance usage privacy& 
Reduced accuracy in utility
\\
\hline
\cite{Won2016}& 
Fault Tolerance&  
Differential Privacy with Encryption (Modular addition) & 
Differential privacy using Laplacian noise with current and future cipher text for fault tolerance with modular additive encryption&
Computationally Complex,
No privacy for individuals
\\
\hline

\cite{Hale2019}& 
Analysis of Accuracy vs Privacy&
Differential Privacy& 
Finding balance at individual level privacy with increased data points for decrease in billing error&

Reduced accuracy in utility
\\
\hline

\cite{sandberg2015differentially}& 
Privacy with State Estimation&  
Differential Privacy& 
Analysis of State estimation vs individual Privacy using differential privacy& 

Lack of analysis on the impact of differential noise on billing
\\
\hline
\end{tabularx}
\label{tab1}
\end{table*}

In the past, proposals by Eibl et al.~\cite{Eibl2017} and Won et al. \cite{Won2016} focus on providing privacy on aggregated data where differentialy perturbed noise is added at trusted aggregator level, protecting user's privacy in the aggregated data. For example, if adversary knows the aggregated data, it can not deduce sensitive information from it. The problem with this approach, is that privatizing aggregated data does not guarantee complete privacy of individuals as unprotected non private aggregated smart meter data can still reveal some critical information about the users \cite{Hale2019}.
%In the light of these services, we subdivide our main goals as follows:
%\begin{itemize}
%\item Preserve individual user's privacy.
%\item Enable aggregator to calculate total energy consumption of all smart meters in an area at an instant in time $t$.
%\item Enable aggregator to calculate total energy consumption of individual users over a period of time $T$.
%\end{itemize}
%In the past, different proposals by Eibl et al \cite{Eibl2017} and Won et al \cite{Won2016} focus on providing privacy on aggregated data where differentialy perturbed noise is added at trusted aggregator level, protecting user's privacy in the aggregated data, for example, if adversary knows the aggregated data, it can not deduce sensitive information from it. An unprotected, non private aggregated smart meter data can reveal some critical information about the users participating in the smart grid. However, privatizing aggregated data does not guarantee complete privacy of individuals.
% Smart meters send data to a third party aggregator for load monitoring and calculation of bills over a period of time. The trusted aggregator adds noise to the sensitive data to provide privacy. However, differential privacy for aggregated data at a trusted third party aggregator level does not provide privacy at individual level.
% In contrast, we provide privacy at an individual level even when the aggregator is untrustworthy. 
To address this challenge Hassan et al.\cite{Hassan2020,Hassan2021} introduced the Differentially Private Dynamic Pricing for Demand Response (DRDP) scheme, providing individual level privacy. In this scheme, the smart meters send original data to the trusted aggregator which masks the data using distributed noise and reports the data to the utility grid along with the billing information. The trusted aggregator stores and calculates the bill according to the original data. The challenge with DRDP, is that it assumes the aggregator as a trusted entity, which introduces significant security risks.

Given the
above context in this paper, we propose a Differential Privacy
with Noise Cancellation Technique (DPNCT) scheme. The main contributions of our DPNCT scheme are as follows: (a) it preserves consumers privacy without using a trusted third
party aggregator, (b) it enables calculation of accurate bills using periodic noise cancellation technique, and (c) it enables accurate load
monitoring using aggregate noise cancellation at aggregator
end. As part of our analysis we have benchmarked DPNCT against DRDP \cite{Hassan2020} with different noise cancellation schemes (hourly, daily, and weekly) on total power consumption at an instant $t$ for load monitoring and total consumption of an individual over a period of time $T$.

% organization of paper
The rest of the paper is organized as follows. Section \ref{litreview} discuss the related work and how our solution differs from them. In section \ref{PS}, we present our proposed solution along with algorithm and example. In section \ref{PA}, we discuss the performance analysis of our scheme and finally conclude the discussion in section \ref{Con}.  

\section{Literature Review}\label{litreview}
Table \ref{tab1} gives an overview of the comparison of different privacy solutions for smart grid using DP. \cite{Eibl2017, Won2016} provides privacy for the aggregated data only using infinite divisibility of Laplacian distribution. As previously mentioned, the challenge with these approaches is that protected aggregated data still can leak useful information regarding individuals. In order to address this issue, Acs et al. \cite{dream} use cryptography schemes, which relies on users sharing cryptographic keys or $ciphertexts$ with each other, which is difficult to manage as the systems scales. Won et al. \cite{Won2016} builds upon the solution provided by \cite{dream} to address the scalability issue and provides fault tolerance by introducing modular additive encryption. Using this approach, smart meters send private data with current and future $ciphertexts$ to cater for future transmission failure, helping system to run smoothly even in scenarios when smart meter fails to share its $ciphertext$. The challenge with the solutions outlined above is that even though they provide DP, their implementation makes them computationally complex and expensive. The most relevant work in smart grid privacy using purely differential privacy is \cite{Barbosa2016, Hale2019, Hassan2020} where they used Laplacian distribution for generation of noise for individual level privacy.  Barbosa et al. \cite{Barbosa2016} used filtering time series attack to validate appliance usage privacy of individual consumers. Trajectory level privacy technique is used by Hale et al. \cite{Hale2019} which protects sensitive smart meter data over a period of time at an individual level and analyze the cost of privacy over accuracy in billing and aggregated load monitoring. By not using a trusted third party, \cite{Barbosa2016} and \cite{Hale2019} introduce a certain level of inaccuracy in bills for the users as a cost of privacy. The authors from \cite{Hassan2020} provide usage based dynamic billing along with differential privacy at aggregator level. The noise is generated at the aggregator level and then added to individual data points before sending it to the grid utility. For dynamic billing, the aggregated load is compared with peak allowed load and only the individuals who are responsible for peak load are charge. However, they depend on a trusted third party and assume a “curious but honest” aggregator to provide privacy. In contrast, in our approach we do not make this assumption, and instead we provide individual level privacy at the smart meter level, before it reaches the aggregator component. In addition, our solution also includes a noise cancellation technique to deal with the error in dynamic billing and load monitoring.  

\begin{figure*}[t]
\centerline{\includegraphics[scale=0.15]{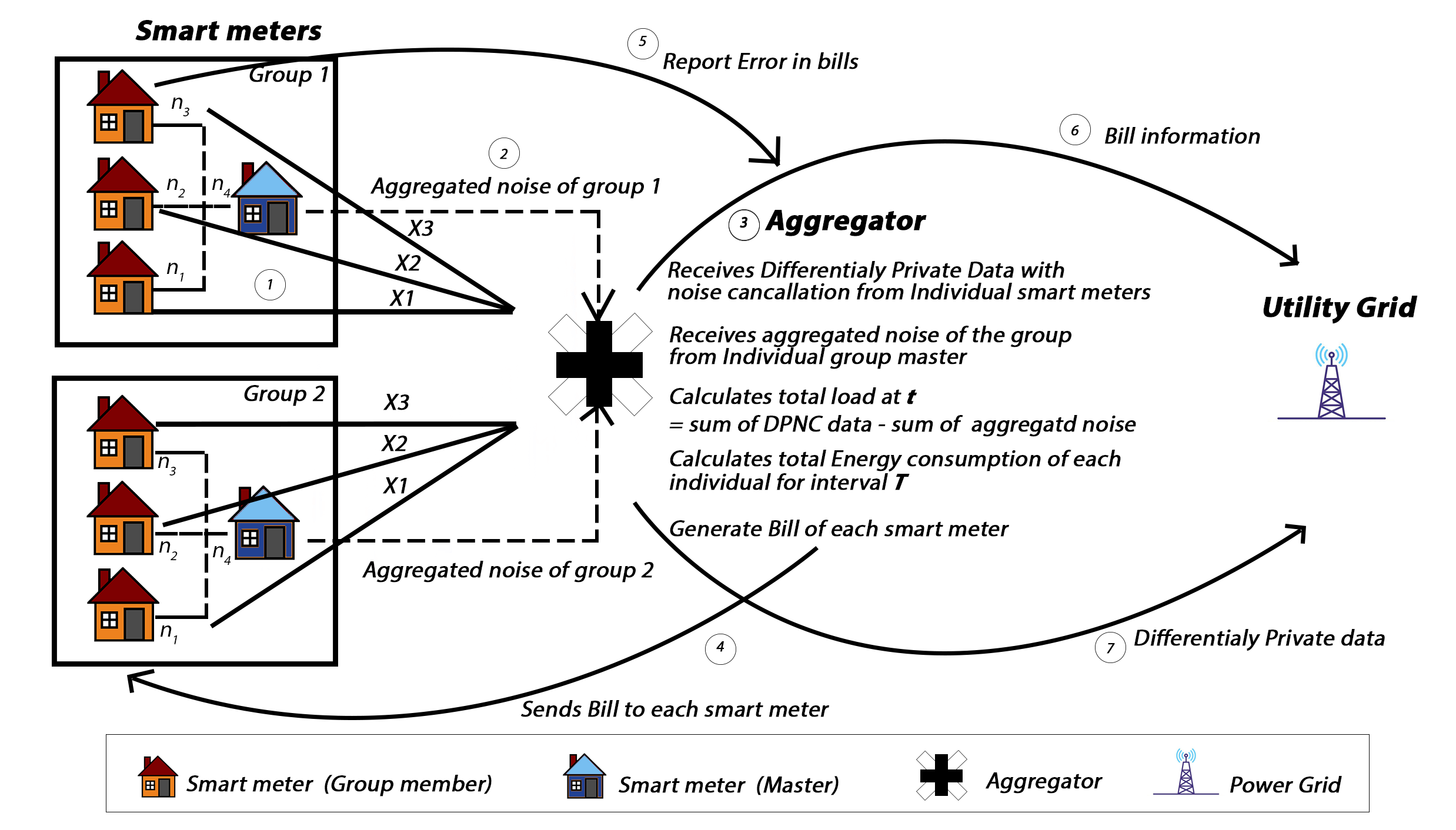}}
\caption{System Model: All smart meters send DPNCT masked data to aggregator and group master sends aggregated noise of the group to the aggregator which is subtracted from total masked data to get true aggregated load.}
\label{sm}
\end{figure*}

\section{Proposed Solution: DPNCT}\label{PS}
In this section, we present our novel solution along with preliminary information of DP as privacy preserving technique.   
\subsection{System Model}

Our model illustrated in Fig.~\ref{sm}, shows three main physical entities: smart meters, aggregators, and utility grid. To calculate total energy consumption in an area at an instant $t$, the aggregator receives differentialy private energy consumption data of each user transmitted by smart meters. However, this data alone does not provide accurate information of total load at an instant $t$ due to the addition of noise in the data at smart meter level. To deal with this issue, smart meters send their added noise at time $t$ to a randomly selected master smart meter shown as blue house in the Fig.~\ref{sm}. The master smart meter then accumulates this noise from all member smart meters in the group and sends it to the aggregator where this accumulated noise gets subtracted from aggregated private data. This process gives an accurate information of total energy consumption at an instant $t$ for load monitoring. 
% If the total energy consumption at time $t$ is greater than the maximum load limit provided by the grid utility, then the aggregator charges the users responsible for this peak load by charging them with peak price per unit $(kWh)$. The aggregator then notify this peak surcharge to the respective smart meters. 
To calculate total energy consumption of an individual, we propose a noise cancelling DP mechanism (DPNCT), where in addition to adding Laplacian noise $n_{\Delta t}$ in time period $\Delta t$, each user subtracts the noise $nc$ added in $\Delta {t-1}$. As a demand response scheme, aggregator checks if the total energy consumption of a single user is more then $maximum Allowed Units$ set by grid utility, then  instead of base unit price, aggregator charges surcharge price for the extra units. However, if the user gets surcharged price due to added noise then the error is corrected in the next bill. \par

\begin{table}[t]
\centering
\caption{Key Notations and their Description used in Algorithm \ref{alg:DPNC1} and \ref{alg:DPNC2}}
% \begin{tabularx}{\textwidth}{|m{2em}|m{10em}|m{2em}|m{10em}|}
\begin{tabularx}{0.5\textwidth}{|c|X|c|X|}
\hline
\textbf{Key}& 
\textbf{Description}& 
\textbf{Key}& 
\textbf{Description}\\
\hline
 $mIDs$ & 
 IDs of all master smart meters &  
$Er_{T-1}$ & 
 Error in previous bill reported by smart meters \\ 
 \hline
 $N$ & 
 Total number of smart meters & 
 $N_K$ & 
Aggregated group noise \\ 
\hline
$x_t$ & 
Original load consumption of the household at time $t$ &  
$\Delta t$ & 
chosen scheme in DPNCT (Hourly, Daily, Monthly) \\ 
\hline
\end{tabularx}
\label{tab3}
\end{table}

% \begin{table}[htbp]
% \centering
% \vspace{0.25cm}
% \caption{Comparison of DPNCT with  DRDP}
% \begin{tabularx}{0.4\textwidth}{|X|c|c|}
% \hline
% \textbf{Feature}& 
% \textbf{DRDP\cite{Hassan2020}}& 
% \textbf{DPNCT}\\
% \hline
% Aggregator-to-grid anonymity & 
% Yes &  
% Yes \\
% \hline
% Dynamic Billing & 
% Yes &  
% Yes \\ 
% \hline
% User-to-Aggregator anonymity & 
% No &  
% Yes \\ 
% \hline
% \end{tabularx}
% \label{tab2}
% \end{table}

\begin{algorithm}
\small
\SetAlgoLined
\SetKwFunction{FnB}{Function BillCalculation()}
\SetKwFunction{FnL}{Function AggregatedLoadCalculation()}

\FnL \;
\Begin {
\KwIn{ $mIDs$} 
% \tcc{$mIDs$ is the set of $K$ selected masters to collect group noise. $Error_{T-1}$ is the Error in bills reported by smart meters.}
% \KwOut{ $$}
% \tcc{$X_N$ is the set of all readings from $N$ smart meters. $N$ is the total number of smart meters in the area.}
 \While{Billing Period $T$}{
 \For{all smart meters $i$ in $N$}
 {
  $X_i$ = getMaskedData($i$)\;
 
  }
  \For{$masterID$ in $mIDs$}{
  $N_K$ = getNoiseData($masterID$)\;
  }
  $totalLoad_{t}$ = $\sum_{i=1}^{N} X_i - \sum_{i=1}^{K} N_i$ \;
 }
 }
 \FnB \;
\Begin {
\KwIn{ $maxUnits, SurchargePrice, UnitPrice$,$Er_{T-1}$}
 \For{ all smart meters $i$ in $N$}{
 \eIf{$\sum_{i}^{T} X_i \geq maxAllaowedUnits$}{
 
 $surchargeUnits = \sum X_i - maxAllaowedUnits$\;
  
 $BaseBill$ = $maxAllaowedUnits * UnitPrice $\;
 $SurchargeBill$ = $surchargeUnits *SurchargePrice$\;
 $TotalBill_i$ =$BaseBill + SurchargeBill - Er_{T-1}$\;
 Notify $TotalBill_i$ and $surchargeUnits$ to smart meter $i$\;
 
 }{
 $TotalBill_i$ = $\sum X_i * UnitPrice$\;
 Notify $TotalBill_i$ to smart meter $i$ \;
 }
 }
 }

 \caption{Calculation of Bill and Aggregated Load at Aggregator}
 \label{alg:DPNC1}
\end{algorithm}

\begin{algorithm}
\small
\SetAlgoLined
\SetKwFunction{FnNC}{Function DPNCT()}
\FnNC \;
\Begin {
\KwIn{$x_t,ID,\Delta t,masterID_t,totalBill,surchargeUnits$}
$N_{t-1}$ = $N_{t}$\;
$N_{t}$ = 0\;
 \While{Time Period $\Delta t$}{
 $n_t$ = G(N,$\lambda$) - G'(N,$\lambda$)\;
 $N_t$ = Push($n_t$)\;
 $nc_{t-1}$ = Pop($N_{t-1}$)\;
 $X_t$ = $x_t$ + $n_t$ - $nc_{t-1}$\;
 Send $X_t$ to aggregator \;
 \eIf{$masterID_t = ID$ }{
 \For{all $k$ smart meters in group}{
 get noise $n_{k,t}$ from member smart meter\;
 }
 Report aggregated group noise $ \sum_k n_{k,t}$ to aggregator
 }{
 Send $n_t$ to master smart meter with $masterID_t$
 }
 }
 \eIf{surcharge Reported By Aggrgator}{
 
 \eIf{$ SurchargeUnits  \geq Total Noise in \Delta t$}{
    Error = $Total Noise$
   
 }{
 Error  = $SurchargeUnits$ \;
 }
  Notify Error To Aggregator \;
 }{
 Error  = $0$\;
 }
}
\caption{Differential Privacy With Noise Cancellation at Smart Meter}
 \label{alg:DPNC2}
\end{algorithm}

\vspace{-0.15cm}
\subsection{Differential Privacy}
As proposed by Dwork et al. \cite{dwork2014algorithmic} differential private noise gives $\epsilon$ privacy for a mechanism, $M$, if for any two neighbouring data-sets $D1$ and $D2$ which differ in at most one record and for all possible answers $S \subseteq Range (M)$, the following equation holds true. 
\begin{equation}\label{Eq:1}
Pr(M(D1) \in  S) \leq e^\epsilon * Pr (M(D2) \in S)
\end{equation}
In simpler terms, it is unlikely that an adversary finds out anything meaningful from smart meters data-set that is differentially private where $\epsilon$ is the privacy parameter controlled by user ranges from 0 to 1. The lesser the value of $\epsilon$ the more private the data would be but, with less utility. 
\subsubsection{Sensitivity}
Sensitivity of a function $f$ is defined as maximum difference in output of any two neighbouring datasets. In our case, we can make use of pointwise sensitivity, explained in detail by Eibl and Engel \cite{Eibl2017}, where each data smart meter $i$ at time $t$ generates noise $n_{i,t}$ independently irrespective of the data of other smart meters. 
\begin{equation}\label{Eq:2}
S_{pw} = max_{D1,D2} | f(D1) - f(D2) | = max_{i,t} |x_{i,t}|
\end{equation}
So the query at time $t$ is $\epsilon_t  = \epsilon / t $ private such that $ \sum \epsilon_t =  \epsilon$ where sensitivity for the data would be maximum consumption by any smart meter at all time. Selection and analysis of different sensitivity strategies is out of scope of this paper. 

\subsubsection{Infinite divisibility of Laplace distribution}
For the privacy of individual consumer, we need to add noise at each smart meter before reporting the data to the aggregator. We use Laplacian noise due to its property of infinite divisibility as each smart meter will add noise on their own independently without any prior knowledge of other smart meters.
Infinite divisibility of Laplace distribution~\cite{dream,Eibl2017} states that if a random variable is sampled from the probability distribution function of Laplace distribution that is: $ f(x,\lambda) = 1/2 (e^{|x|/\lambda})$, then the distribution is infinitely distributed for $N \geq 1$, 
\begin{equation}\label{Eq:3}
Lap(\lambda) = \sum_{i=1}^{N} (G(N,\lambda) - G'(N,\lambda))
\end{equation}
Where $G$ and $G'$ are independent and identical distributed gamma density functions with same parameters. $N$ is the number of smart meters at network level and $\lambda$ is drawn on the basis of $\epsilon$ and point wise sensitivity. 
Equation \ref{Eq:3} implies that at an instant $t$, the aggregated noise of all smart meters would be equal to $Lap(\lambda)$ when using gamma density function. 

\begin{figure*}[t]
\begin{subfigure}[t]{.5\textwidth}
  \centering
  \includegraphics[width=1\linewidth]{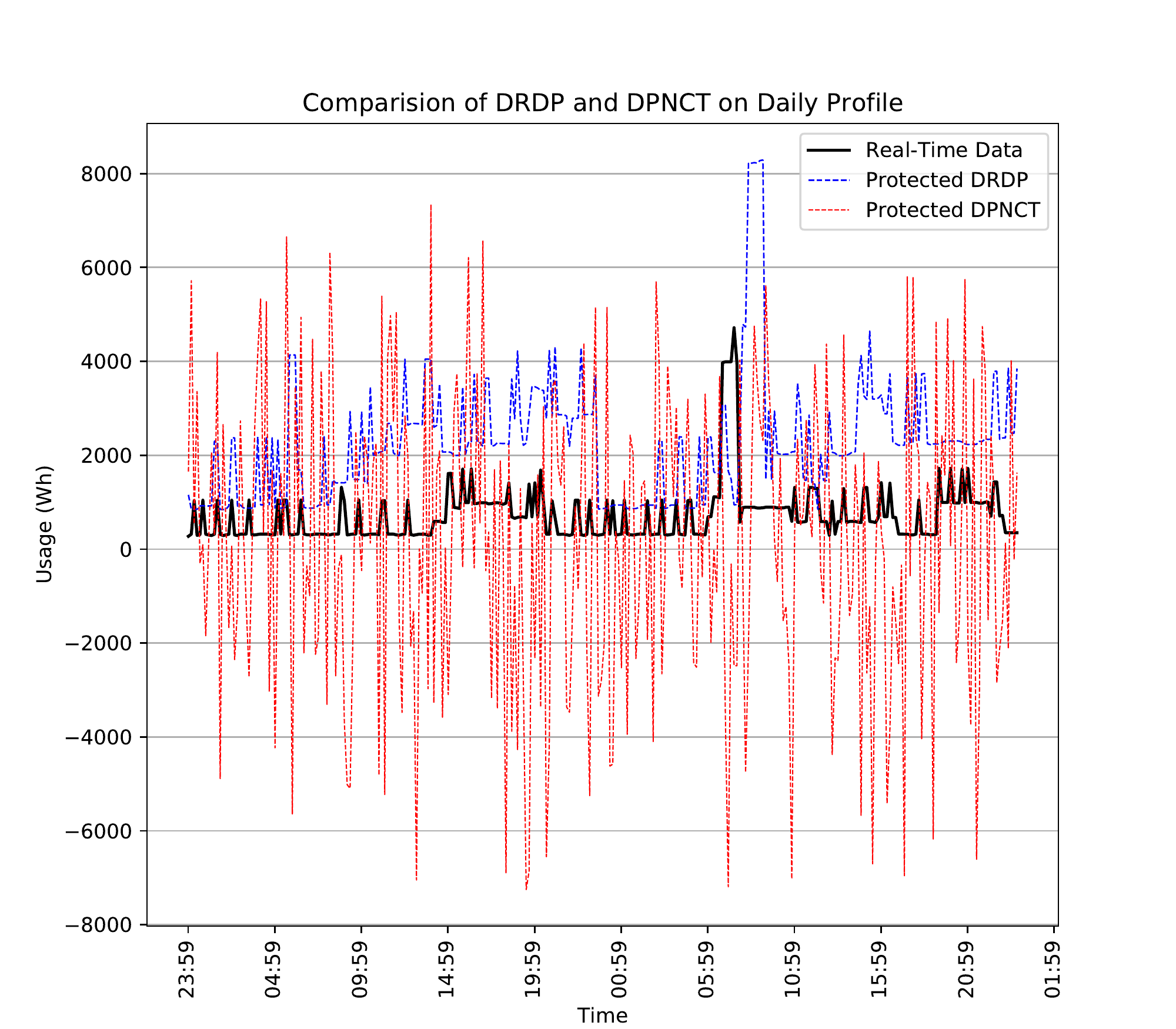}
    \caption{}%Comparison of DRDP and DPNCT with real-time data. This graph shows that the daily usage is well masked by both DRDP and DPNCT schemes. But with DPNCT, the masking is much better.}
 \label{fig:smuneeb1}
\end{subfigure}
\begin{subfigure}[t]{.5\textwidth}
  \centering
  \includegraphics[width=1\linewidth]{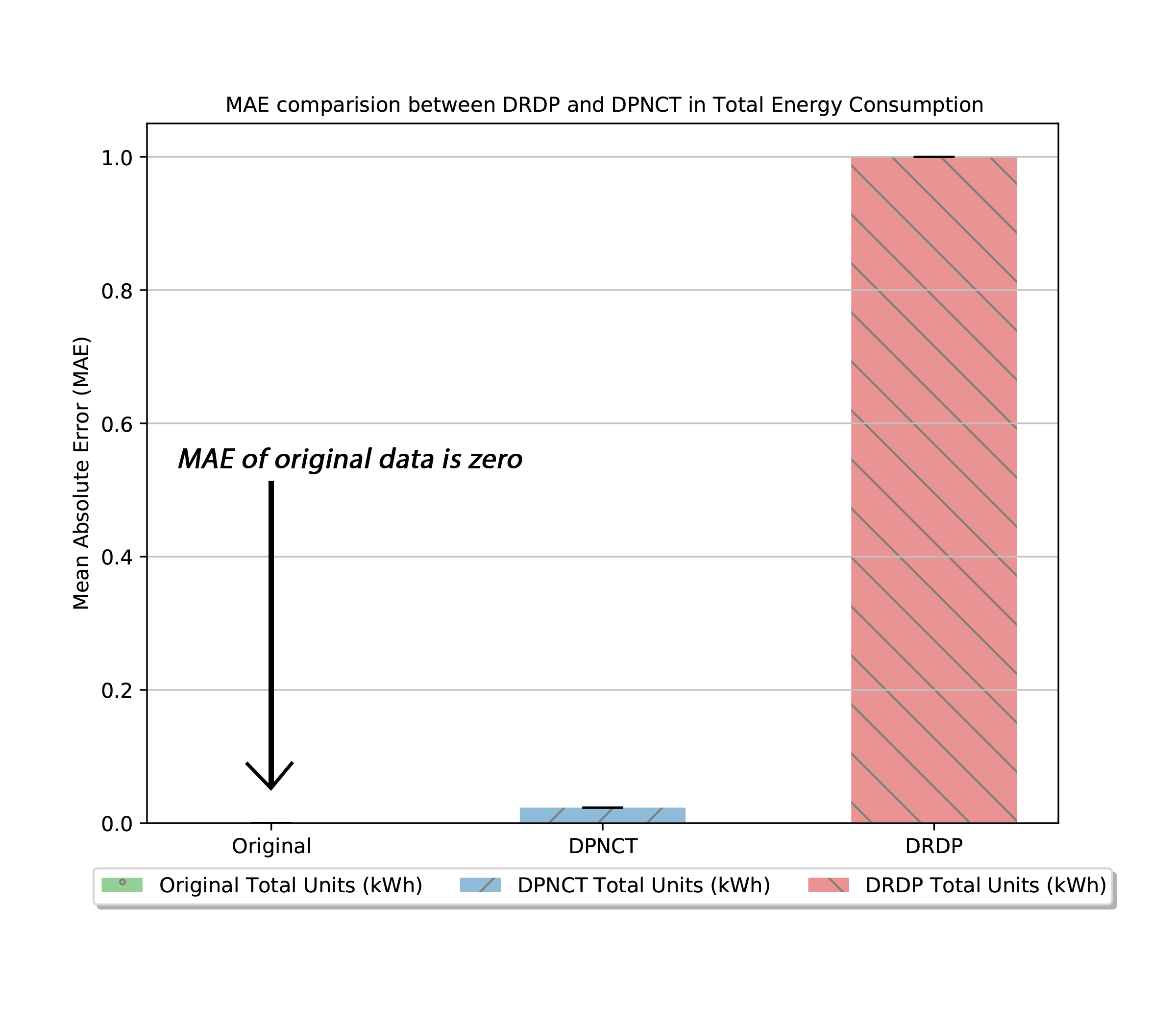}
    \caption{}%Comparison of MAE between Original Unprotected data, DRDP, and DPNCT in total energy consumption.}
  \label{fig:smuneeb2}
\end{subfigure}
\label{fig:muneeb}
\caption{Evaluation of differential privacy and comparison of DRDP and DPNCT with real-time daily profile of a randomly selected house. Fig. 2(a) shows comparison of DRDP and DPNCT with real-time data. This graph shows that the daily usage is well masked by both DRDP and DPNCT schemes. But with DPNCT, the masking is much better. Fig. 2(b) shows comparison of MAE between Original Unprotected data, DRDP, and DPNCT in total energy consumption.}
\end{figure*}

\subsection{Differentially Private Noise Cancellation Mechanism}  
We assume that our smart grid model has $N$ smart meters and one aggregator. Each smart meter $i$ records its power consumption reading $x_{i,t}$ in $kWh$ at an instant $t$. Since, aggregator does not need to know the individual consumption of users, each smart meter $i$ adds gamma noise to its original energy consumption data at time $t$ i.e. $ x_{i,t} + (G(N,\lambda) - G'(N,\lambda)) $ and sends this masked data to the aggregator. Using equation \ref{Eq:3}, the masked data $X_{i,t}$ of $N$ smart meters gives differential privacy of $\epsilon$ when aggregated as follows.  
\begin{equation}
\sum_{i=1}^{N} x_{i,t} + (G(N,\lambda) - G'(N,\lambda)) = \sum_{i=1}^{N} x_{i,t} + Lap(\lambda) = \sum_{i=1}^{N} X_{i,t}
\end{equation}
However, to increase the accuracy of aggregated load at an instant $t$, we use aggregated noise cancellation protocol. In this protocol, each smart meter is assigned an $ID$ by aggregator and in each round $K$ groups are formed. Each group has $k$ out of $N$ smart meters randomly selected. A master $k_i$ is selected randomly in each group and all members send their noise to the master which then send the aggregated group noise to the aggregator. Using function
AggregatedLoadCalculation from Algorithm \ref{alg:DPNC1}, the aggregator subtract the aggregated group noise i.e., $\sum_{i=1}^{k} n_{i,t}$ from total masked values ($X_{i,t}$) to get accurate load at time $t$ as follows. 
\begin{equation}
\sum_{i=1}^{n} X_{i,t} - \sum_{i=1}^{k} n_{i,t} = \sum_{i=1}^{n} x_{i,t} 
\end{equation}
To improve accuracy in billing, each smart meter records noise added to the smart meter data over a period of time $\Delta t$. Each smart meter generate gamma noise $n_{i,t}$ independently using equation \ref{Eq:3} and adds it to the original data before reporting to the aggregator. The total noise added in $\Delta t$ is subtracted from the smart meter data in the next period of time $\Delta t$ to cancel the overall effect of noise in billing. We will see the effect of selecting this time period $\Delta t$ schemes in performance evaluation section.
The protocol is further elaborated in function DPNCT Algorithm \ref{alg:DPNC2} with the help of Table \ref{tab3}.

\section{Performance Analysis}\label{PA}

In this section, we evaluate our algorithm for privacy and accuracy. The experiments are performed over the energy consumption data provided by \cite{data} and results are compared with the benchmark set by DRDP \cite{Hassan2020}. In \cite{data}, residential energy profiles in watts of 200 households with the granularity of 10 minutes is provided which gives $T = 6 * 24 * 30 = 4,320$ data points per month for a single household. For implementation of DPNCT, we used Numpy library of  Python 3.0 (cf. https://numpy.org). For simplicity, we used $\epsilon = 1$ and point-wise sensitivity $max_{i,t} |x_{i,t}|$ with $mean = 0$ to calculate scale parameter $\lambda$ for Laplacian noise generation. The complexity cost of generating a random number is $O(1)$ and our algorithm adds a random number i.e., noise $n_t$ at each reading $x_{i,t}$ so the complexity of our Algorithm per smart meter is $O(N)$, $N$ being the total number of data points in time period $T$ same as DRDP \cite{Hassan2020}. For noise cancellation, we keep track of the  noise added in previous period $\Delta t_{t-1}$ and the same noise is then subtracted in the next period $\Delta t_{t}$. We compare noise cancelling schemes with $\Delta t$ as hourly, daily, and weekly. 
For dynamic billing we set $Max Allowed Units$ to be $5500kWh$ and $Unit$ and $SurchargePrice$ to be $10\$$ and $20\$$ respectively. 
All the experiments were performed 20 times and the average of them is taken as to normalise the nature of randomness in the noise cancellation and noise generation.\par

% we used the Dirichlet Distribution for randomized generation of noise. We choose the Dirichlet Distribution because of its support to generate $N$ random variables such that they sums up to 1. We then multiply the distribution with the total noise and get a distribution that sums to exactly the noise added in previous period.

In the Fig. 2, we compare our DPNCT Technique  with the results of DRDP strategy used by \cite{Hassan2020} on the daily profile of a randomly chosen single user.
In the given Fig. \ref{fig:smuneeb1}, the solid black line denotes original real-time data and the dotted blue line shows protected data by DRDP, the dotted red line depicts DPNCT protected data. The masking effect of noise added by DPNCT technique has close to none correlation ($0.11$, $1$ being the highest correlation) with the real-time data profile. 
This low correlation depicted in Fig. \ref{fig:smuneeb1}, means that an adversary cannot infer a users behaviour and life style patterns, ensuring the privacy of user data patterns generated without the underlying assumption of a trusted third party aggregator.\par

\begin{table}[t]
\centering
\vspace{0.25cm}
\caption{Comparison of DPNCT with  DRDP}
\begin{tabularx}{0.4\textwidth}{|X|c|c|}
\hline
\textbf{Feature}& 
\textbf{DRDP\cite{Hassan2020}}& 
\textbf{DPNCT}\\
\hline
Aggregator-to-grid anonymity & 
Yes &  
Yes \\
\hline
Dynamic Billing & 
Yes &  
Yes \\ 
\hline
User-to-Aggregator anonymity & 
No &  
Yes \\ 
\hline
\end{tabularx}
\label{tab2}
\end{table}

As demonstrated in the Table \ref{tab2}, our proposed DPNCT, ensures user-to-aggregator anonymity as an additional feature over DRDP. We calculated Mean Absolute Error (MAE) in total energy consumption of a single household as follows~\cite{dream}:
\begin{equation}\label{Eq:4}
MAE = \sum_{i=1}^{N} \frac {|x_i - X_i|}{x_i}
\end{equation}
Where $|x_i - X_i|$ is the absolute error between sum of real values and total DPNCT masked values of a household. 
In Fig. \ref{fig:smuneeb2}, we compare MAE in total energy consumption of a single household between DPNCT hourly scheme and DRDP. 
The impact of DPNCT schemes on the utility goals of smart metering data i.e., billing and load aggregation for load monitoring and dynamic pricing, is analysed below.

%\vspace{-0.25cm}

\subsubsection{Billing}

\begin{figure}[t]
  \centering
  \includegraphics[width=1\linewidth]{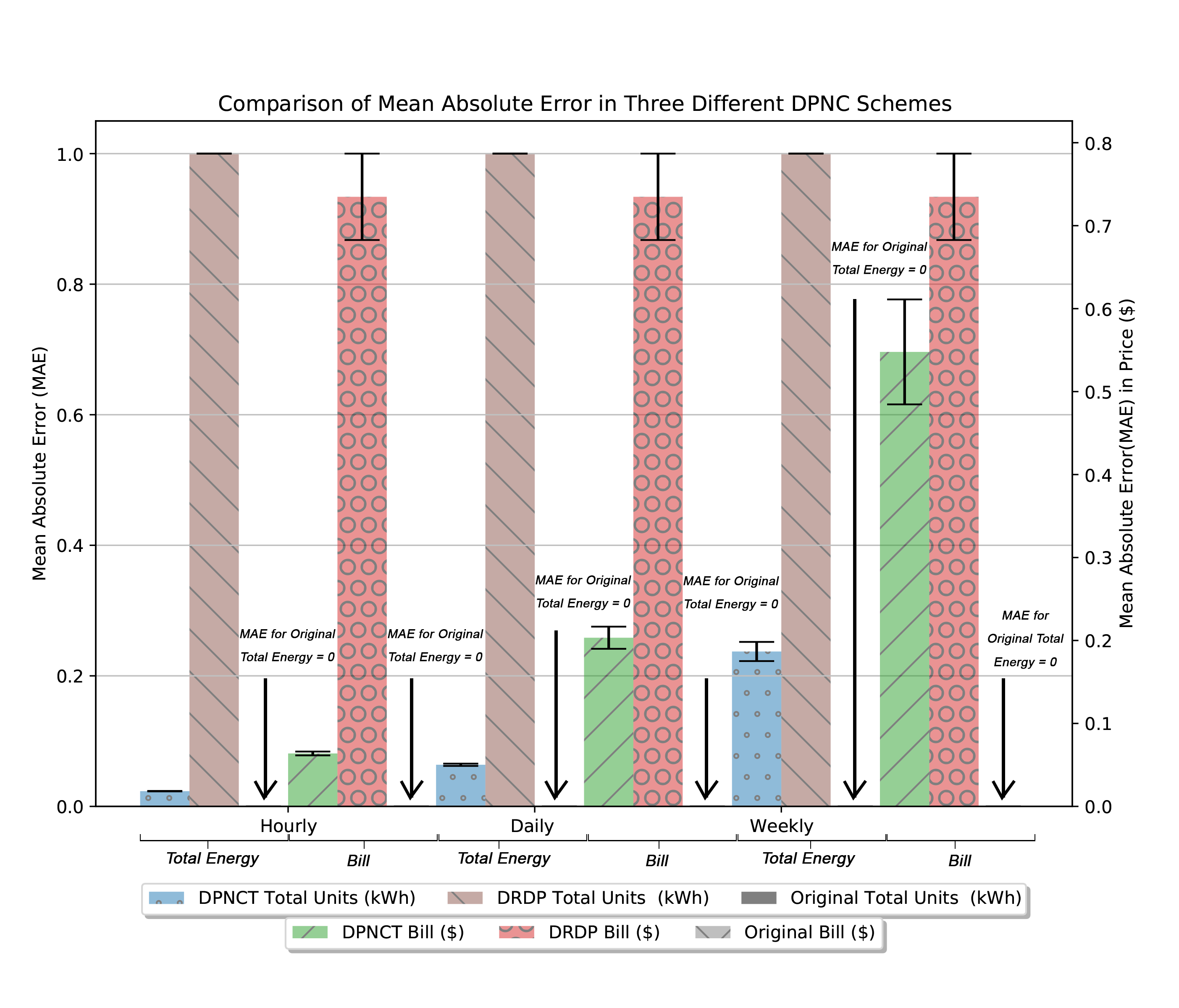}
  \caption{Comparison of Mean Absolute Error in different schemes of DPNCT for total consumption and dynamic bill of a randomly selected household. }
  \label{fig:error}
\end{figure}

For billing period $T$, if a single meter $i$ with energy consumption $x_{i,t}$ provides the $\epsilon_{i}$ differential privacy at an instant $t$ then the total error in the bill would be noise $Lap(\lambda)_{\Delta t}$ added in the last $\Delta t $ of the billing period $T$, where $\Delta t $ can be an hour or a day or a week, according to the selected noise cancellation technique. 
% We set the peak aggregated load to $12000 Wh$ and used data of $10$ house holds for experiments. On average each house is allowed $1200 Wh$ at an instant $t$ in a cluster of 10 houses. If the user consumes more than the average allowance at peak time then the user is charged with peak factor price per unit. In our Algorithm \ref{alg:DPNC2}, if the added noise $n_{i,t}$ in $X_{i,t}$ is the cause of peak factor pricing then the noise $n_{i,t}$ is multiplied by the same peak factor and the effect of surcharge is deducted in the next  $\Delta t$. \par
In Fig. \ref{fig:error}, we compare different noise cancellation period schemes i.e., hourly, daily, and weekly. We calculated Mean Absolute Error (MAE) in total energy consumption ($kWh$) of a arbitrarily selected single household. In Fig. \ref{fig:error},  we also compared the effect of different schemes on our dynamic billing scheme. The MAE in hourly noise cancellation scheme for total energy consumption was the lowest ($0.045$) because of the least amount of noise left at the end of the billing period. For example, in hourly noise cancelling scheme, if a total noise $n_{t1}$ of $7kWh$ is added in the hour 12:00 - 01:00 then the cancelling noise of exact same amount i.e., $7kWh$ is subtracted in the next hour 01:00 - 02:00. 
The MAE at the end of billing period for hourly noise cancellation scheme was the lowest ($0.06$) because the bill only has small error added due to the addition of noise in the last hour of last day of the billing period.
The MAE  in total energy consumption of daily and weekly schemes are $0.2$ and $0.5$ respectively. As the error in bill is reported to the aggregator and it gets corrected in the next billing period, the customer sees no impact in terms of billing given the operation of the DPNCT Algorithm \ref{alg:DPNC1}. 
% Whereas the MAE percentage in total bill of daily and weekly schemes are $27\%$ and $33\%$ respectively. 

\subsubsection{Load Monitoring}
For Load Monitoring at an instant $t$, %each $x_{i,t}$ provides the $\epsilon_{t}$ at instant $t$ then the total privacy would be $\sum \epsilon_t $. 
in best case scenario, the average error in aggregated load would be zero due to aggregated noise cancellation as all the $k$ groups send aggregated noise at an instant $t$. However, in worst case scenario where no accumulated noise would be reported by any group then the total noise at an instant $t$ would be $Lap(\lambda)$. This means the worst case scenario can be improved by selecting robust value for sensitivity instead of overall maximum. %Different statistical techniques are used by \cite{Won2016,Hale2019} to increase the utility of aggregated load, which is one of our future goals. %however, improvement and analysis of worst case aggregated load error is included in our future goals. 

\section{Conclusion }\label{Con}
In this paper, we proposed a privacy preserving solution for smart meters with maximum utility for bill calculation and aggregated load monitoring using noise cancellation technique. Further, we cancel the effect of noise on the surcharge billed to the customer due to the added noise. In this way, minimizing the financial impact of privacy on the customer while preserving the privacy. DPNCT provides less MAE in total energy consumption and in billing as compared to DRDP without trusted third party.
%DPNCT provides $5\%$ MAE in total energy consumption and $21\%$ in billing as compared to DRDP which provides $100\%$ MAE in total energy consumption and $75\%$ in billing. 
% We further conclude that the hourly noise cancelling scheme preforms best as it only gives $5\%$ MAE in total energy consumption of single household as compared to $13\%$ and $37\%$ MAE in daily and weekly schemes respectively.
Similarly, privacy at the individual level precludes the requirement of a trusted third party and ensures that adversary will not be able to deduce users' life style and sensitive behavioural information from collected data. In future, we will work on the selection of sensitivity and analysis of its impact on aggregated load monitoring as well as analysis of trust model for master smart
meters.%\vspace{-0.2cm} %We will extend this work in combination with different DR techniques.

\section*{Acknowledgement}
\vspace{-0.2cm}This publication has emanated from research conducted with the financial support of Science Foundation Ireland (SFI) and is funded under the Grant Number 18/CRT/6222.\vspace{-0.2cm}

\medskip

%Sets the bibliography style to UNSRT and imports the 
%bibliography file "samples.bib".
\bibliographystyle{IEEEtran}
\bibliography{DPwithNC}

\begin{thebibliography}{10}
\providecommand{\url}[1]{#1}
\csname url@samestyle\endcsname
\providecommand{\newblock}{\relax}
\providecommand{\bibinfo}[2]{#2}
\providecommand{\BIBentrySTDinterwordspacing}{\spaceskip=0pt\relax}
\providecommand{\BIBentryALTinterwordstretchfactor}{4}
\providecommand{\BIBentryALTinterwordspacing}{\spaceskip=\fontdimen2\font plus
\BIBentryALTinterwordstretchfactor\fontdimen3\font minus
  \fontdimen4\font\relax}
\providecommand{\BIBforeignlanguage}[2]{{%
\expandafter\ifx\csname l@#1\endcsname\relax
\typeout{** WARNING: IEEEtran.bst: No hyphenation pattern has been}%
\typeout{** loaded for the language `#1'. Using the pattern for}%
\typeout{** the default language instead.}%
\else
\language=\csname l@#1\endcsname
\fi
#2}}
\providecommand{\BIBdecl}{\relax}
\BIBdecl

\bibitem{inferPrivLeak}
M.~A. {Lisovich}, D.~K. {Mulligan}, and S.~B. {Wicker}, ``Inferring personal
  information from demand-response systems,'' \emph{in IEEE Security Privacy},
  vol.~8, no.~1, pp. 11--20, 2010.

\bibitem{privInvationUmass}
A.~Molina-Markham, P.~Shenoy, K.~Fu, E.~Cecchet, and D.~Irwin, ``Private
  memoirs of a smart meter,'' in \emph{Proceedings of the 2nd ACM Workshop on
  Embedded Sensing Systems for Energy-Efficiency in Building}, ser. BuildSys
  '10, New York, NY, USA, 2010, p. 61–66.

\bibitem{dwork2014algorithmic}
C.~Dwork, A.~Roth \emph{et~al.}, ``The algorithmic foundations of differential
  privacy.'' \emph{Foundations and Trends in Theoretical Computer Science},
  vol.~9, no. 3-4, pp. 211--407, 2014.

\bibitem{Hassan2020Comst}
M.~U. {Hassan}, M.~H. {Rehmani}, and J.~{Chen}, ``Differential privacy
  techniques for cyber physical systems: A survey,'' \emph{IEEE Communications
  Surveys \& Tutorials}, vol.~22, no.~1, pp. 746--789, 2020.

\bibitem{dream}
G.~Ács and C.~Castelluccia, ``I have a dream! (differentially private smart
  metering),'' vol. 6958 LNCS, 2011.

\bibitem{Eibl2017}
G.~Eibl and D.~Engel, ``Differential privacy for real smart metering data,''
  \emph{Comput Sci Res Dev}, vol.~32, p. 173–182, 2017.

\bibitem{Hassan2020}
M.~U. {Hassan}, M.~H. {Rehmani}, and J.~{Chen}, ``Differentially private
  dynamic pricing for efficient demand response in smart grid,'' in \emph{IEEE
  International Conference on Communications (ICC)}, 2020, pp. 1--6.

\bibitem{Hassan2021}
\BIBentryALTinterwordspacing
------, ``Differentially private demand side management for incentivized
  dynamic pricing in smart grid.'' 2021. [Online]. Available:
  \url{https://arxiv.org/abs/2102.01478}
\BIBentrySTDinterwordspacing

\bibitem{Barbosa2016}
P.~Barbosa, A.~Brito, and H.~Almeida, ``A technique to provide differential
  privacy for appliance usage in smart metering,'' \emph{Information Sciences},
  vol. 370-371, 2016.

\bibitem{Won2016}
J.~{Won}, C.~Y.~T. {Ma}, D.~K.~Y. {Yau}, and N.~S.~V. {Rao}, ``Privacy-assured
  aggregation protocol for smart metering: A proactive fault-tolerant
  approach,'' \emph{IEEE/ACM Transactions on Networking}, vol.~24, no.~3, pp.
  1661--1674, June 2016.

\bibitem{Hale2019}
M.~Hale, P.~Barooah, K.~Parker, and K.~Yazdani, ``Differentially private smart
  metering: Implementation, analytics, and billing,'' in \emph{Proceedings of
  the 1st ACM International Workshop on Urban Building Energy Sensing,
  Controls, Big Data Analysis, and Visualization}, ser. UrbSys'19, New York,
  NY, USA, 2019, p. 33–42.

\bibitem{sandberg2015differentially}
H.~Sandberg, G.~D{\'a}n, and R.~Thobaben, ``Differentially private state
  estimation in distribution networks with smart meters,'' in \emph{2015 54th
  IEEE conference on decision and control (CDC)}, 2015, pp. 4492--4498.

\bibitem{data}
M.~Muratori, ``Impact of uncoordinated plug-in electric vehicle charging on
  residential power demand-supplementary data.'' \emph{National Renewable
  Energy Laboratory-Data, Golden, CO (United States)}, 2017.

\end{thebibliography}
% \bibliography{anystyle}

\vspace{12pt}

\end{document}